\documentclass[aps,prl,twocolumn,groupedaddress,amsmath,superscriptaddress]{revtex4-1}

\usepackage{amsfonts}
\usepackage{amsmath}
\usepackage[dvips]{graphicx}
\usepackage{color}
\usepackage{amsmath}
\usepackage{graphicx}
\usepackage{amssymb}
\usepackage{hyperref}
\usepackage{color}
\usepackage{mathrsfs}
\usepackage{isomath}
\usepackage{amsmath}
\usepackage{amsthm}
\usepackage{epstopdf}
\usepackage{txfonts}
\usepackage{dsfont}
\usepackage{ulem}
\usepackage{float}
\allowdisplaybreaks[4]

\makeatletter

\definecolor{nblue}{rgb}{0.3,0.3,1.0}
\definecolor{ngreen}{rgb}{0.2,0.7,0.2}
\definecolor{nred}{rgb}{0.9,0.1,0}
\definecolor{nblack}{rgb}{0,0,0}

\newcommand{\beq}{\begin{equation}}
    \newcommand{\eeq}{\end{equation}}
\newcommand{\bqa}{\begin{eqnarray}}
	\newcommand{\eqa}{\end{eqnarray}}

\makeatletter

\usepackage{babel}

\begin{document}

\title{Experimental demonstration of remotely creating Wigner negativity via quantum steering}

\author{Shuheng Liu$^{\ddagger}$}
\affiliation{State Key Laboratory for Mesoscopic Physics, School of Physics, Frontiers Science Center for Nano-optoelectronics, Peking University, Beijing 100871, China}
\author{Dongmei Han$^{\ddagger}$}
\affiliation{State Key Laboratory of Quantum Optics and Quantum Optics Devices, Institute of Opto-Electronics, Shanxi University, Taiyuan, 030006, China}
\affiliation{Collaborative Innovation Center of Extreme Optics, Shanxi University,
Taiyuan, Shanxi 030006, China}
\author{Na Wang}
\affiliation{State Key Laboratory of Quantum Optics and Quantum Optics Devices, Institute of Opto-Electronics, Shanxi University, Taiyuan, 030006, China}
\affiliation{Collaborative Innovation Center of Extreme Optics, Shanxi University,
Taiyuan, Shanxi 030006, China}
\author{Yu Xiang}
\affiliation{State Key Laboratory for Mesoscopic Physics, School of Physics, Frontiers Science Center for Nano-optoelectronics, Peking University, Beijing 100871, China}
\affiliation{Collaborative Innovation Center of Extreme Optics, Shanxi University,
Taiyuan, Shanxi 030006, China}
\author{Fengxiao Sun}
\affiliation{State Key Laboratory for Mesoscopic Physics, School of Physics, Frontiers Science Center for Nano-optoelectronics, Peking University, Beijing 100871, China}
\affiliation{Collaborative Innovation Center of Extreme Optics, Shanxi University,
Taiyuan, Shanxi 030006, China}
\author{Meihong Wang}
\affiliation{State Key Laboratory of Quantum Optics and Quantum Optics Devices, Institute of Opto-Electronics, Shanxi University, Taiyuan, 030006, China}
\affiliation{Collaborative Innovation Center of Extreme Optics, Shanxi University,
Taiyuan, Shanxi 030006, China}
\author{Zhongzhong Qin}
\affiliation{State Key Laboratory of Quantum Optics and Quantum Optics Devices, Institute of Opto-Electronics, Shanxi University, Taiyuan, 030006, China}
\affiliation{Collaborative Innovation Center of Extreme Optics, Shanxi University,
Taiyuan, Shanxi 030006, China}
\author{Qihuang Gong}
\affiliation{State Key Laboratory for Mesoscopic Physics, School of Physics, Frontiers Science Center for Nano-optoelectronics, Peking University, Beijing 100871, China}
\affiliation{Collaborative Innovation Center of Extreme Optics, Shanxi University,
Taiyuan, Shanxi 030006, China}
\affiliation{Peking University Yangtze Delta Institute of Optoelectronics, Nantong, Jiangsu, China}
\author{Xiaolong Su}
\email{suxl@sxu.edu.cn}
\affiliation{State Key Laboratory of Quantum Optics and Quantum Optics Devices, Institute of Opto-Electronics, Shanxi University, Taiyuan, 030006, China}
\affiliation{Collaborative Innovation Center of Extreme Optics, Shanxi University,
Taiyuan, Shanxi 030006, China}
\author{Qiongyi He}
\email{qiongyihe@pku.edu.cn}
\affiliation{State Key Laboratory for Mesoscopic Physics, School of Physics, Frontiers Science Center for Nano-optoelectronics, Peking University, Beijing 100871, China}
\affiliation{Collaborative Innovation Center of Extreme Optics, Shanxi University,
Taiyuan, Shanxi 030006, China}
\affiliation{Peking University Yangtze Delta Institute of Optoelectronics, Nantong, Jiangsu, China}

\begin{abstract}

Non-Gaussian states with Wigner negativity are of particular interest in quantum technology due to their potential applications in quantum computing and quantum metrology. However, how to create such states at a remote location remains a challenge, which is important for efficiently distributing quantum resource between distant nodes in a network. Here, we experimentally prepare optical non-Gaussian state with negative Wigner function at a remote node via local non-Gaussian operation and shared Gaussian entangled state existing quantum steering. By performing photon subtraction on one mode, Wigner negativity is created in the remote target mode. We show that the Wigner negativity is sensitive to loss on the target mode, but robust to loss on the mode performing photon subtraction. This experiment confirms the connection between the remotely created Wigner negativity and quantum steering. As an application, we present that the generated non-Gaussian state exhibits metrological power in quantum phase estimation. 

\end{abstract}

\maketitle

Generation and manipulation of quantum states are crucial preconditions underlying various quantum information tasks. Gaussian states which can be generated deterministically have been widely applied in continuous variable (CV) quantum information~\cite{Samuel2005,Christian2012,wangxb2007,suxl2010,jingjt2020}.
On the other hand, CV non-Gaussian states are attracting increasing interests due to the increasing entanglement contributed from higher-order correlations~\cite{akira2006,alexei2007,takahashi2010entanglement}, and especially, the Wigner negativity of the non-Gaussian states has been identified as an essential resource for reaching a quantum computation advantage~\cite{menicucci2006universal,mari2012} and for error correction~\cite{lund2008error,Victor2018}. Substantial progress has been made in controllable generation of Wigner-negative states by performing non-Gaussian operations, e.g., photon addition or subtraction on the previously prepared Gaussian modes~\cite{Wenger04,Zava04,Ourjoumtsev2006,Neergaard2006,Valentina2007,kentaro2007,nicolas2020}, or directly by a higher-order interaction such as three-photon spontaneous parametric down-conversion~\cite{Douady2004,Chang2020} or four-wave mixing with Kerr nonlinearity~\cite{Leghtas2015}. 

Beyond above local preparation methods, remote state preparation (RSP) based on the shared entanglement between distant nodes offers intrinsic security and efficiency for creating desired quantum resources at a remote location~\cite{Lo2000remote,Paris2003continuous,pogorzalek2019secure,optica2018}. Compared to the well-known quantum teleportation, which transmits an unknown state by sharing entanglement and performing joint Bell measurement, RSP protocol only requires measurements acting on each individual mode. This promises RSP various potential applications in quantum information processing, such as on-demand preparation of single-photon states~\cite{jeffery2004towards,peters2005remote}, creating two-qubit hybrid entangled states~\cite{barreiro2010remote}, generating and manipulating atomic quantum memories remotely~\cite{rosenfeld2007remote,bao2012quantum}. 
Toward networked quantum technology it is, thus, crucial to find a way to prepare a remote  quantum state with a negative Wigner function.

Recently, it has been theoretically shown that a special kind of entanglement known as Einstein-Podolsky-Rosen (EPR) steering~\cite{Schrodinger35,wiseman2007,reid2009,cava2017,Uola2020} is a necessary requirement for remotely preparing Wigner-negative states~\cite{mattiaPRL,mattiaPRXquantum,XiangarXiv}. EPR steering is a directional form of nonlocality, related to the Einstein ``spooky" paradox, that after performing local measurements on one of the systems, it can apparently steer the state of the other distant system. Based on this kind of nonlocal effect existing between distant systems, one can remotely create a Wigner-negative state in the steering mode by subtracting a photon from the steered mode~\cite{mattiaPRL}. At the same time, the shared entanglement still maintains as the photon subtraction is a non-destructive non-Gaussian operation. This connection has not been experimentally verified yet, especially the quantitative relation when taking the practical channel loss between spatially separated nodes into account. Thus, it is still quite an open area for further investigations.

In this Letter, we experimentally demonstrate remote creation of non-Gaussian state with negative Wigner function via local single-photon operation and shared Gaussian EPR steering. Two optical modes $A$ and $B$ in a CV EPR entangled state are sent to two distant stations controlled by Alice and Bob respectively  [Fig.~\ref{figExperimentalSchematic}(a)]. Once Alice successfully subtracts a photon from the steered mode $A$, the Wigner function of the steering mode $B$ shows negative values. We quantify the Wigner negativity by performing quantum tomography on the conditional state of mode $B$, and validate the relation between the initially shared Gaussian EPR steering and the remotely created Wigner negativity. The dependence of Wigner negativity on channel loss is investigated by transmitting Alice's and Bob's states through lossy channels, respectively. The results show that the generated Wigner negativity is sensitive to the loss in Bob's channel but robust to the loss in Alice's channel. As an application, we show that the generated non-Gaussian state exhibits metrological power in quantum phase estimation. Our work demonstrates the feasibility of remote preparation of Wigner-negative state between spatially separated stations, and confirms the connection between remotely created Wigner negativity and quantum steering.
\begin{figure}[tb]
\centering
\includegraphics[width=0.9\linewidth]{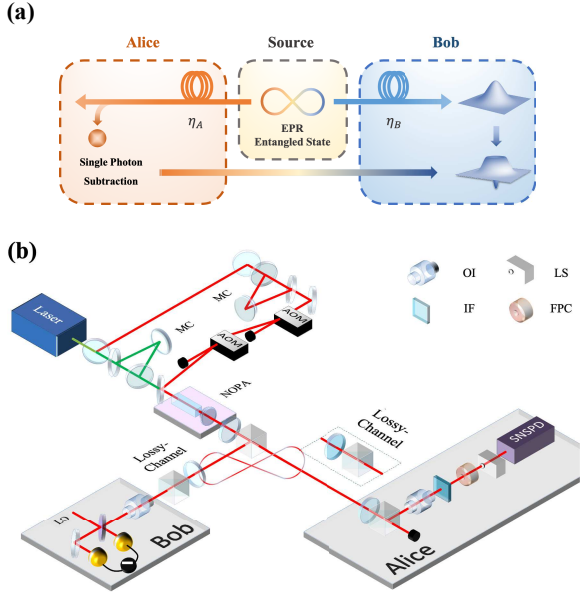}
\caption{The principle and experimental setup. (a) Schematic of the remote preparation of Wigner negativity. We first prepare a Gaussian EPR entangled state and then transmit two entangled optical fields to two distant nodes controlled by Alice and Bob, where the lossy channels are characterized by $\eta_{A}$ and $\eta_{B}$, respectively. Then once Alice successfully performs a single-photon subtraction from her mode, the remote Bob's mode collapses to a Wigner-negative state. (b) Experimental setup. Two acousto-optic modulators (AOM) controlled by the periodically signals are used to chop the seed beam. The NOPA is composed of a type-II KTP crystal and a concave mirror with 50 mm radius. Lossy channel is simulated by the combination of a half wave plate (HWP) and a polarization beamsplitter (PBS). The optical isolators are used to avoid the back scattered light to the NOPA cavity. SNSPD: superconducting nanowire single-photon detector, LO: local oscillator, MC: mode cleaner, OI: optical isolator, LS: laser shutter, IF: interference filter, FPC: Fabry-Perot cavity.}
\label{figExperimentalSchematic}
\end{figure}

The experimental setup is shown in Fig.~\ref{figExperimentalSchematic}(b). A continuous laser generates $1080$ nm and $540$ nm laser beams simultaneously, which are used as the seed and pump beams of a nondegenerate optical parametric amplifier (NOPA). An EPR entangled state is generated from the NOPA when it is operated at the status of deamplification~\cite{liu2019,suddensu}. Two modes of the EPR entangled  state are separated by a polarization beam splitter (PBS), and transmitted to Alice and Bob through lossy channels, characterized by $\eta_{A}$ and $\eta_{B}$, respectively. Alice then performs single-photon subtraction by splitting her mode with a beamsplitter with around $4\%$ reflectivity and implementing single-photon detection on it. The filter system used to select the degenerate mode is composed of an interference filter with 0.6 nm bandwidth and a Fabry-Perot cavity (FPC). When a photon is detected by the superconducting nanowire single-photon detector (SNSPD), Bob measures his conditional state with a homodyne detector. 
\begin{figure*}[tb]
\centering
\includegraphics[width=0.85\linewidth]{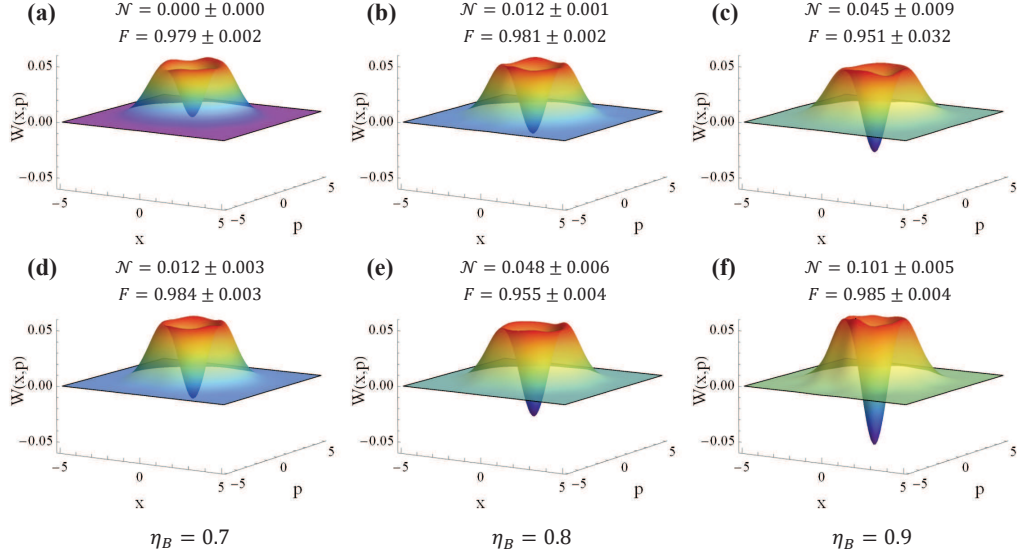}
\caption{The reconstructed Wigner function of Bob's state remotely created by performing single-photon subtraction from Alice's field with two sets of input squeezing levels: (a-c) $-$1.74/+2.08 $\text{dB}$ and (d-f) $-$1.302/+1.407 $\text{dB}$, under different transmission efficiencies of mode $B$: $\eta_B=0.7,~0.8,~0.9$. Note that in our experiment, $10\%$ detection loss is caused by the limited homodyne detection efficiency including the quantum efficiency of the photo diode ($98\%$), the mode matching efficiency ($98\%$), and the clearance of homodyne detection ($96\%$), which is corrected in the above plot.}
\label{figWignerFunctions}
\end{figure*}

The CV EPR entangled state shared between Alice and Bob can be described by its covariance matrix with elements $\sigma _{ij}=\langle \hat{\beta}_{i}\hat{\beta}_{j}+\hat{\beta}_{j}
\hat{\beta}_{i}\rangle /2-\langle \hat{\beta}_{i}\rangle \langle \hat{\beta}%
_{j}\rangle $, where $\hat{\beta}\equiv (\hat{x}_A,\hat{p}_A,\hat{x}_B,\hat{p}_B)^{\top}$ is the vector of the amplitude and phase
quadratures of each mode. The quadrature operators of Alice's mode $A$ are denoted by $\hat{x}_{A}=\hat{a}_{A}^{\dagger}+\hat{a}_{A}, \hat{p}_{A}=i\left(\hat{a}_{A}^{\dagger}-\hat{a}_{A}\right)$ where $\hat{a}^{\dagger}, \hat{a}$ are creation and annihilation operators respectively. Same definition for Bob's mode $B$. Thus the covariance matrix is given by 
\begin{equation}
\sigma_{A B}=
\left(\begin{array}{cc}\sigma_{A} & \gamma_{A B} \\ \gamma_{A B}^{\top} & \sigma_{B}\end{array}\right)
=\left(\begin{array}{cccc}
n & 0 & c_{1} & 0 \\ 
0 & n & 0 & c_{2} \\ 
c_{1} & 0 & m & 0 \\ 
0 & c_{2} & 0 & m 
\end{array}\right),
\end{equation}
where the submatrices $\sigma_A$ and $\sigma_B$ represent the statistical features of the reduced states of subsystems $A$ and $B$, respectively; $\gamma_{AB}$ provides cross correlations between the output optical modes. In our experiment, the realized CV EPR resources with $c_1=-c_2=c$. The CM elements of Eq. (1) can be retrieved from single mode measurements, {\it i.e.}, by simultaneously measuring the amplitude and phase quadratures of each of two output modes~\cite{JOB, PRA69, PRA76, PRA87, OL46,EPJ05,PRL09}. 

EPR steering in the direction from Bob to Alice is quantified by the parameter $\mathcal{G}^{B\rightarrow A}=\max\{0, \frac12\ln\frac{\text{Det~}\sigma_{B}}{\text{Det~}\sigma_{AB}}\}$~\cite{Adesso15}, a higher value means stronger steerability. Note that $\mathcal{G}^{B\rightarrow A}>0$ has been theoretically proved as a sufficient and necessary resource to remotely generate negative part of Wigner function of the steering mode $B$ by a conditional operation applied on the steered mode $A$~\cite{mattiaPRL}. 

To confirm this connection, we then subtract a single photon from the steered mode $A$ and observe its nonlocal effects on the steering mode $B$. After a successful single-photon subtraction, the Wigner function of the reduced quantum state of mode $B$ is expressed as
\begin{equation}
\label{eqWignerFunction}
\begin{aligned}
W^{A-}_{B}(\beta_{B})=&\frac{\exp \left\{-\frac{1}{2}\left(\beta_{B}, \sigma_{B}^{-1} \beta_{B}\right)\right\}}{2 \pi \sqrt{\operatorname{Det} \sigma_{B}}\left[\operatorname{Tr}\left(\sigma_{A}\right)-2\right]}\\
&\times \left[\beta_{B}^{\top} \sigma_{B}^{-1^{\top}} \gamma_{A B}^{\top}  \gamma_{A B} \sigma_{B}^{-1} \beta_{B}+ \operatorname{Tr}(V_{A \mid B})-2\right],
\end{aligned}
\end{equation}
where $\beta_{B}=(x_B,~p_B)^{\top}$ is the vector of possible measurement outcomes of the quadrature operators, and $V_{A \mid B}=\sigma_{A}-\gamma_{A B} \sigma_{B}^{-1} \gamma_{A B}^{\top}$ is the Schur complement of $\sigma_{A}$. The non-classical features of the reduced quantum state can be characterized by the negativity of the above Wigner function, defined as the doubled volume of the integrated negative part~\cite{kenfack2004negativity}, 
\begin{equation}
\mathcal{N}_{B}^{A-}=\frac{2 c^{2} e^{\frac{m(n-1)}{c^{2}}-1}}{m(n-1)}-2.
\end{equation}
The derivation is detailed in ~\cite{supp}, where we also express the created Wigner negativity as a function of the local and global purities of initial Gaussian EPR entangled state before photon subtraction, confirming the quantitative relation given in a recent theoretical work~\cite{XiangarXiv}.
\begin{figure}[t]
	\centering
	\includegraphics[width=\linewidth]{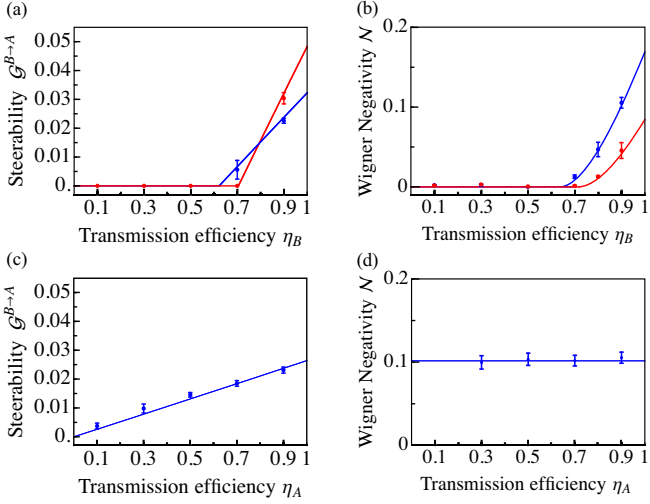}
	\caption{The steerability $\mathcal{G}^{B\rightarrow A}$ of the initial Gaussian entangled states and the remotely created Wigner negativity in the steering mode $B$ after subtracting a single photon from the steered mode $A$ as functions of the transmission efficiency $\eta_{B}$ shown in (a,~b), or $\eta_{A}$ in (c,~d), respectively. In (a,~b), we also examine two sets of input squeezing levels, $-$1.74/+2.08 $\text{dB}$ (red lines) and $-$1.302/+1.407 $\text{dB}$ (blue lines). Error bars represent $\pm$ one standard deviation and are obtained based on the statistics of the measured noise variances and density matrices.}
\label{figWignerNegativity}
\end{figure}

When taking the dark counts of the single-photon detector into account, the Wigner function given in Eq.~(\ref{eqWignerFunction}) is then changed to $W_{B,\text{s}}^{A-}=\xi W_{B}^{A-}+(1-\xi) W_{B}$~\cite{darkcount}, where $\xi$ is the ratio of the true counts from single-photon subtraction on $A$ to the total detector clicks, $W_{B}^{A-}$ is the ideal Wigner function of photon-subtracted state and $W_{B}$ is the initial Gaussian state which corresponds to the failure event of single-photon subtraction. Thus, the negativity $\mathcal{N}_{B,\text{s}}^{A-}$ of Wigner function $W_{B,\text{s}}^{A-}$ when considering dark counts reads
\begin{equation}
\mathcal{N}_{B,\text{s}}^{A-}=
\frac{2 c^{2} \xi e^{\frac{m(n-1)}{c^{2}\xi}-1}}{m(n-1)}-2.
\end{equation}
The experimental details of the click rate (generation rate) and $\xi$ can be found in~\cite{supp}.

In the experiment, we first prepare a Gaussian entangled state from a NOPA and retrieve its covariance matrix from single mode measurements~\cite{supp}. In particular, the amplitude and phase quadratures of optical fields $A$ and $B$ are measured by homodyne detectors in the time domain, where the signals of detectors pass through two low-pass filters with bandwidth of $60$ MHz and are recorded simultaneously by a digital storage oscilloscope at the sampling rate of $500$ KS/s. Two different entangled Gaussian states characterized by different squeezing levels are generated at the source by injecting $50$ mW and $30$ mW pump beams into the NOPA respectively, which results in different purities after a lossy evolution from the crystal downward the detection~\cite{ra1,ra2,ra3}.

To investigate the effects of squeezing level and purity on the remote creation of Wigner negativity, we analyze the elements of the measured CM in terms of $V_{\pm}=\Delta^2(\hat{x}_A\pm\hat{x}_B)/2=\Delta^2(\hat{p}_A\mp\hat{p}_B)/2$, which are the correlated variances of the quadrature measurement statistics between two modes of the EPR entangled states~\cite{kurochkin2014PRL}. Thus, considering the practical transmission efficiency, the CM elements of Eq. (1) can be expressed as $n=\eta_A(V_++V_-)/2+(1-\eta_A)$, $m=\eta_B(V_++V_-)/2+(1-\eta_B)$, $c_{1}=-c_{2}=c=-\sqrt{\eta_A\eta_B}(V_--V_+)/2$~\cite{supp}. In our experiment, the entangled state generated by $30$ mW pump power shows lower squeezing but higher purity $-1.302/+1.407$ dB ({\it i.e.}, $V_+=0.74,~V_-=1.38$), while by $50$ mW pump power has higher squeezing but lower purity $-1.74/+2.08$ dB ({\it i.e.}, $V_+=0.67,~V_-=1.61$). 

To perform the single-photon subtraction on the steered mode $A$, around $4\%$ energy of optical field $A$ is reflected by a beamsplitter and directed to SNSPD. When the SNSPD clicks at Alice's station, Bob measures his conditional state with a homodyne detector and record the output signals by the digital storage oscilloscope. Note that in case of no click at SNSPD, the non-Gaussian operation fails and the Bob’s conditional state remains Gaussian. We record over $30000$ quadrature values of Bob's mode for each chosen transmission efficiency, then reconstruct the Wigner functions of mode $B$ by using the maximum-likelihood algorithm~\cite{lvovsky2009}.  

Fig.~\ref{figWignerFunctions} shows the reconstructed Wigner functions of mode $B$ conditioned on single-photon subtraction performed on the distant mode $A$ at different transmission efficiencies $\eta_B$ and fixed $\eta_A=0.9$ for two sets of input squeezing levels: $-$1.74/+2.08 $\text{dB}$ (a-c) and $-$1.302/+1.407 $\text{dB}$ (d-f). The corresponding Wigner negativities $\mathcal{N}$ become larger with the increase of $\eta_{B}$. The fidelity $F(\rho,\sigma)=\left(\operatorname{Tr}\sqrt{\sqrt{\rho}\sigma\sqrt{\rho}}\right)^2$ is a measure which quantifies the overlap between the experimentally reconstructed reduced quantum state of mode $B$ ($\sigma $) after single-photon subtraction on mode $A$ and the theoretical result with $\xi$. They are all above $95\%$ for the presented transmission efficiencies $\eta_B$, which manifests high quality of the RSP process. The results reveal that the case with lower squeezing $-$1.302/+1.407 $\text{dB}$ indicated in Figs.~\ref{figWignerFunctions}(d-f) performs better than the other case with $-$1.74/+2.08 $\text{dB}$ given in Figs.~\ref{figWignerFunctions}(a-c), showing more significant negative values as the transmission efficiency increases. Especially, for the transmission efficiency $\eta_B=0.7$, Fig.~\ref{figWignerFunctions}(d) already presents nonzero Wigner negativity, while Fig.~\ref{figWignerFunctions}(a) doesn't.

Since this remote preparation of non-Gaussian state of mode $B$ is based on Gaussian EPR steering shared between modes $A$ and $B$, to understand the physics behind the results given by Fig.~\ref{figWignerFunctions}, we investigate the connection between the Wigner negativity and the Gaussian steerability from a quantitative perspective, and establish the decisiveness factors imposing constraints on the degree of Wigner negativity that is remotely created by single-photon subtraction performed in a distant station. Especially, we take into account the lossy channels and examine the effects of loss in the protocol. 

As shown in Fig.~\ref{figWignerNegativity}(a), Gaussian steerability from Bob to Alice only exists ($\mathcal{G}^{B\rightarrow A}>0$) when $\eta_{B}>0.623$ (the case with lower squeezing but higher purity described in blue) or $\eta_{B}>0.701$ (the case with higher squeezing but lower purity indicated in red) for a fixed value of $\eta_A=0.9$. Correspondingly, the Wigner negativity of the reduced state of mode $B$ appears only when Gaussian steerability is larger than zero, as shown in Fig.~\ref{figWignerNegativity}(b). The states produced with a higher squeezing level possess stronger Gaussian steerability as expected, while the created Wigner negativity in the steering mode $B$ is lower (red lines). This means there is a one-to-one correspondence of nonzero Gaussian steerability and Wigner negativity, but lack of a quantitative connection between them. In fact, there is a tradeoff between the quality of the initial EPR entanglement (the two-mode squeezing) and the final Wigner negativity. For instance, if the entanglement (squeezing) is too high the reduced single-mode state (by tracing over the mode that was subject to photon subtraction) has a ``higher temperature"~\cite{lvovsky2015squeezed} which prevents (high) Wigner negativity after photon subtraction. As indicated in~\cite{supp}, the state purity plays main role instead of the squeezing level in our experiment.
\begin{figure}[t]
\centering
\includegraphics[width=0.95\linewidth]{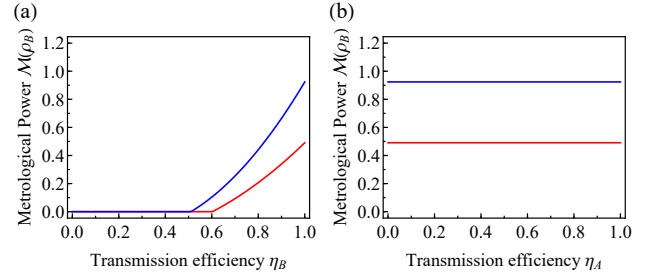}
\caption{The metrological power remotely created in mode $B$ after single-photon subtraction on mode $A$, varying with (a) transmission efficiency $\eta_B$ of Bob and (b) transmission efficiency $\eta_A$ of Alice. The squeezing levels are $-$1/+1 $\text{dB}$ (blue) and $-$3/+3$\text{dB}$ (red)}.
\label{figMetrologicalPower}
\end{figure}

The Gaussian steerability enhances with the increase of transmission efficiency in two channels, as shown in Figs.~\ref{figWignerNegativity}(a) and (c), while the created Wigner negativity is only affected by the channel loss existing in the steering mode $B$ (b), but does not vary with the loss in the channel of the steered mode (d). This is because that Wigner function of the reduced state only depends on $\eta_B$, as shown in Eq.~(\ref{WNB}) in the Appendix~\cite{supp}. However, the transmission efficiency in Alice's channel will affect the generation rate of non-Gaussian state at Bob's node. For instance, for the case with lower squeezing, the generation rate of creating Wigner negativity in mode $B$ is decreased from $\sim$3 kHz to $\sim$500 Hz when $\eta_A$ decreases from $0.9$ to $0.3$. 

As an application of the remotely prepared non-Gaussian state with Wigner negativity, we examine its metrological power in quantum precision measurement, as demonstrated in Fig.~\ref{figMetrologicalPower}. The metrological power is defined as $\mathcal{M}(\rho)=1/4 \max[F_x(\rho)-2,0]$~\cite{metrology2019PRL,metrologyPRR,metrologyPRA}, where $\mathcal{M}(\rho)$ quantifies the metrological advantage beyond the standard quantum limit, and $F_x(\rho)$ is the optimized quantum Fisher information over all possible quadratures $\hat{x}$~\cite{toth2014quantum,demkowicz2015quantum}. Similar with the Wigner negativity, the metrological power of the reduced state becomes stronger with a lower level of input squeezing (blue lines), and sensitive to the loss existing in Bob's channel but robust to the loss in Alice's channel. 

In summary, we experimentally prepare non-Gaussian state with negative Wigner function in a remote mode (hold by Bob) by performing local single-photon subtraction on the steered mode (hold by Alice) of the previously shared Gaussian EPR entangled state. We demonstrate that the appearance of Wigner negativity at Bob's node depends on the presence of the steerability from Bob to Alice, confirming the connection between the remotely created Wigner negativity and quantum steering. To further examine the quantitative relation we take the channel loss between two distant nodes, and find that the created Wigner negativity is sensitive to the loss in channel head to Bob, but robust to the loss in channel head to Alice who performs photon subtraction. We also show a potential application of the prepared non-Gaussian state in quantum phase estimation, where less initial squeezing produces higher Wigner negativity and thus stronger metrological power in the steering mode. Our results present a significant advance in a concrete in-depth understanding of the connection between remotely creating Wigner negativity and the Gaussian EPR steering, and pave the way for remote preparation of multimode non-Gaussian states for reaching further quantum advantage.

\begin{acknowledgments}
This work is supported by National Natural Science Foundation of China (Grants No. 12125402, No. 11834010, No. 12004011), National Key R\&D Program of China (Grant No. 2019YFA0308702). X. S. thanks the Fund for Shanxi ``1331 Project" Key Subjects Construction. Q. H. acknowledges the Beijing Natural Science Foundation (Z190005) and the Key R\&D Program of Guangdong Province (Grant No. 2018B030329001).

\end{acknowledgments}

$^{\ddagger}$S. H. Liu and D. M. Han contributed equally to this work.

\appendix*
\setcounter{equation}{0}
\section*{Appendix I: The derivation of remotely created Wigner negativity}

In our experiment, an EPR entangled state was directly generated from the nondegenerate optical parametric amplifier, which is fully described by its covariance matrix (CM)
\begin{equation}
\sigma_{A B}=
\left(\begin{array}{cc}\sigma_{A} & \gamma_{A B} \\ \gamma_{A B}^{\top} & \sigma_{B}\end{array}\right)
=\left(\begin{array}{cccc}
n & 0 & c_{1} & 0 \\ 
0 & n & 0 & c_{2} \\ 
c_{1} & 0 & m & 0 \\ 
0 & c_{2} & 0 & m 
\end{array}\right),
\label{cm}
\end{equation}
where $n=\Delta^{2}\hat{x}_{A}=\Delta^{2}\hat{p}_{A}$, $m=\Delta^{2}\hat{x}_{B}=\Delta^{2}\hat{p}_{B}$ represent the variances of amplitude and phase quadratures of the output optical modes, $c_{1}=Cov(\hat{x}_{A},\hat{x}_{B})$ and $c_{2}=Cov(\hat{p}_{A},\hat{p}_{B})$ indicate their cross correlations. The CM elements of Eq.~(\ref{cm}) can be retrieved from single mode measurements as detailed in APPENDIX II-B~\cite{JOB, PRA69, PRA76, PRA87, OL46,EPJ05,PRL09}.

We can define the correlated variances of the quadrature measurement statistics between two modes of the EPR entangled states by $V_+=\Delta^2(\hat{x}_A+\hat{x}_B)/2=\Delta^2(\hat{p}_A-\hat{p}_B)/2$ and $V_-=\Delta^2(\hat{x}_A-\hat{x}_B)/2=\Delta^2(\hat{p}_A+\hat{p}_B)/2$.
Considering the practical losses characterized by transmission efficiency $\eta_A$ and $\eta_B$, the CM elements become $n=\eta_A(V_++V_-)/2+(1-\eta_A)$, $m=\eta_B(V_++V_-)/2+(1-\eta_B)$, $c_{1}=-c_{2}=c=-\sqrt{\eta_A\eta_B}(V_--V_+)/2$. The local purities $\mu_{A,B}$ of modes $A,~B$ and the global purity $\mu_{AB}$ are then given by

\begin{align}
&\mu_A=\frac{1}{\sqrt{\text{det} \sigma_A}}=\frac{2}{2+(V_-+V_+-2)\eta_A}, \notag\\
&\mu_B=\frac{1}{\sqrt{\text{det} \sigma_B}}=\frac{2}{2+(V_-+V_+-2)\eta_B},\\
&\mu_{AB}=\frac{1}{\sqrt{\text{det}\sigma_{AB}}} \notag\\
&=\frac{2}{2 \eta_{A} \eta_{B} (V_--1)
   (V_+-1)+\eta_{A} (V_-+V_+-2)+\eta_{B}
   (V_-+V_+-2)+2}. \notag
\end{align}

The Gaussian steerability $\mathcal{G}^{B\rightarrow A}$ in such system can be also quantified from its CM~\cite{AdessoPRL2015}, 
\begin{equation}
\begin{aligned}
&\mathcal{G}^{B\rightarrow A}=\max\left\{0,\frac{1}{2}\ln\frac{\text{det}\sigma_B}{\text{det}\sigma_{AB}}\right\}=\max\left\{0,\ln\frac{\mu_{AB}}{\mu_{B}}\right\}\\
&=\max \left\{0,-\ln \left| \frac{
\left[V_-+V_++2 (V_--1) (V_+-1) \eta_{B}-2\right]\eta_{A}}{(V_-+V_+-2) \eta_{B}+2}+1\right| \right\}.
\end{aligned}
\end{equation}
After a single-photon subtraction applied on the steered mode $A$, the Wigner function of the steering mode $B$ becomes~\cite{mattiaPRL} 
\begin{equation}
\begin{aligned}
&W_{B}^{A-}\left(x_B,p_B\right)=\frac{\exp[{-\frac{x_{B}^2+p_{B}^2}{2 m}}] \left[-2 c^2 m+c^2 \left(x_{B}^2+p_{B}^2\right)+2 m^2 (n-1)\right]}{4 \pi  m^3 (n-1)} \\
&=\frac{\exp[-\frac{x_{B}^2+p_{B}^2}{\eta_{B}
(V_-+V_+-2)+2}]}{\pi (V_-+V_+-2) \left[\eta_{B} (V_-+V_+-2)+2\right]^3} \bigg\{\eta_{B} (x_{B}^2+p_{B}^2) (V_--V_+)^2 \\
& \quad +2 \left[2 \eta_{B} (V_--1) (V_+-1)+V_-+V_+-2\right] \left[\eta_{B} (V_-+V_+-2)+2\right] \bigg\}.
\end{aligned}
\label{WNB}
\end{equation}
It is clearly seen that the above Wigner function only depends on the channel loss existing in the steering mode $B$, i.e., $\eta_B$, but does not vary with the loss in the channel of the steered mode $A$. It is straightforward to calculate the Wigner negativity possessed in mode $B$ which is defined as the doubled volume of the integrated negative part of the above Wigner function~\cite{kenfack2004negativity}
\begin{equation}
\mathcal{N}_{B}^{A-}=\frac{2 c^{2} e^{\frac{m(n-1)}{c^{2}}-1}}{m(n-1)}-2.
\end{equation}
Since the Wigner negativity is calculated based on Eq.~(\ref{WNB}), it should be irrelevant to $\eta_A$. 

When taking dark counts of the single-photon detector into account, we have $W_{B, \text{s}}^{A-}=\xi W_{B}^{A-}+(1-\xi) W_{B}$, where $\xi$ is the ratio of the true counts from single-photon subtraction on $A$ to the total detector clicks, and $W_{B}$ is the initial Gaussian Wigner function of mode $B$ corresponding to the failure event of single-photon subtraction~\cite{darkcount}. Thus, the negativity of the Wigner function $W_{B, \text{s}}^{A-}$ accounting for dark counts is expressed as
\begin{equation}
\mathcal{N}_{B, \text{s}}^{A-}=\frac{2 c^{2} \xi e^{\frac{m(n-1)}{c^{2} \xi}-1}}{m(n-1)}-2.
\end{equation}
We can also express it by the local and global purities of the initial Gaussian EPR entangled state,
\begin{equation}
\label{eqWNPurity}
\mathcal{N}_{B, \text{s}}^{A-}=\frac{2 \xi  (\mu_{A} \mu_{B}-\mu_{AB})
e^{\frac{\mu_{AB}-\mu_{A} \mu_{AB}}{\mu_{AB} \xi -\mu_{A} \mu_{B}
\xi }-1}}{(\mu_{A}-1) \mu_{AB}}-2.
\end{equation}
It is clearly that the Wigner negativity $\mathcal{N}_{B, \mathrm{s}}^{A-}$ is determined by the purities of initial Gaussian state.

\section*{Appendix II: Details of experiment}
\subsection*{A. Generation of EPR entangled state}
\begin{figure*}[tb]
\centering
\includegraphics[width=1\linewidth]{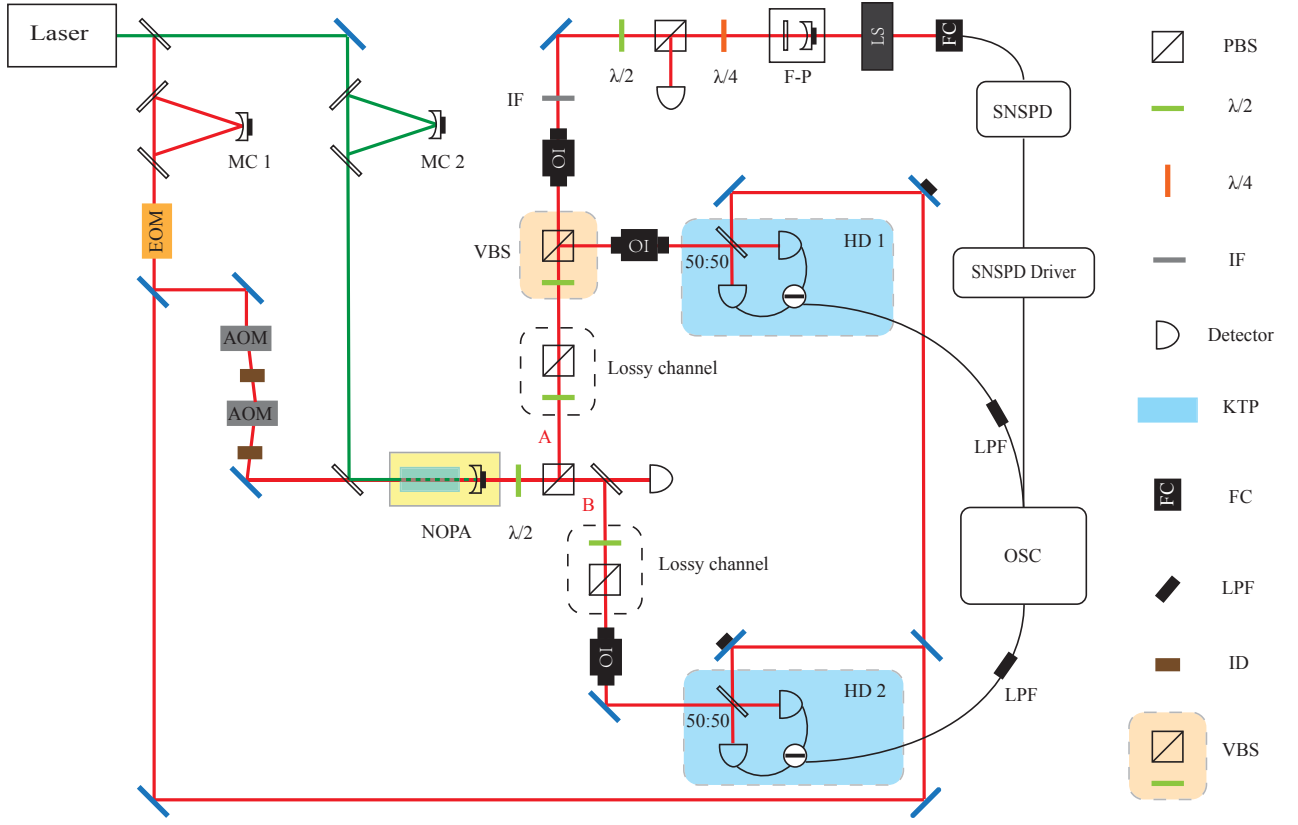}
\caption{Experimental setup. MC 1: mode cleaner for the $1080$ nm light, MC 2: mode cleaner for the $540$ nm light, EOM: electro-optic modulator, AOM: acousto-optic modulator, ID: iris diaphragm, Det: detector, NOPA: non-degenerate optical parametric amplifier, PBS: polarization beam splitter, VBS: variable beam splitter which is composed of a half wave plate (HWP) and a PBS, OI: optical isolator, IF: interference filter, FPC: Fabry-Perot cavity, LS: laser shutters, LPF: low-pass filter, SNSPD: superconducting nanowire single-photon detector, LO: local oscillator, HD 1: homodyne detector at Alice's station, HD 2: homodyne detector at Bob's station, OSC: oscilloscope.}
\label{figExperimentalSchematic}
\end{figure*}

The detailed experimental setup is shown in Fig.~\ref{figExperimentalSchematic}. The laser used in our experiment is a continuous wave intracavity frequency-doubled and frequency-stabilized Nd:YAP-LBO (Nd-doped YAlO$_{\text{3}}$ perovskite-lithium triborate) laser which generates $1080$ nm and $540$ nm light simultaneously. The $1080$ nm and $540$ nm lasers are filtered by mode cleaners (MC 1 and MC 2) respectively and then used as the seed and pump beams of the non-degenerate optical parametric amplifier (NOPA). Two acousto-optic modulators (AOM) are used to chop the seed beam into a cyclic form: $50$ ms for each locking and hold period. During the locking period, the seed beam is injected into the NOPA for the cavity locking. When the seed beam is chopped off, the cavity length of NOPA is holding and the EPR entangled state is generated. The measurement is performed during the hold period.

The NOPA is composed of an $\alpha $-cut type-II KTiOPO4 (KTP) crystal ($3\times3\times10~mm^3$) and a concave mirror with curvature $50$ mm. The front face of the KTP crystal is used as the input coupler, of which the transmittances at $540$ nm and $1080$ nm are $40\%$ and $0.04\%$, respectively. The end face of the KTP crystal is antireflection coated for both $1080$ nm and $540$ nm. The transmittances of the output coupler are $12.5\%$ and $0.5\%$ at $1080$ nm and $540$ nm, respectively. Two modes of the EPR entangled state are separated by a polarization beam splitter (PBS) placed behind the NOPA. 

\subsection*{B. Measurement of Gaussian covariance matrix}

As shown in Fig.~\ref{figExperimentalSchematic}, mode $A$ is totally reflected by a variable beam splitter (VBS) and directed to homodyne detector 1 (HD 1), and mode $B$ is transmitted through a lossy channel simulated by a half wave plate (HWP) and a PBS. The optical isolators (OI) placed before homodyne detectors are used to avoid the back-scattered light from the local oscillators. The output signals of HDs are filtered by two $60$ MHz low-pass filters (LPFs) and recorded by a digital storage oscilloscope (OSC, TELEDNE LECROY, HDO8108A). Quadrature values of two modes in the time domain are sampled with the sampling rate of $500$ MS/s. The relative phase of HD is locked to $0(90)$ degrees to measure the amplitude(phase) quadrature. 

The OI together with mirrors in the signal path to each HD lead to around $10\%$ transmission loss. In addition to this, the detection efficiency of each HD is around $90\%$ which includes the quantum efficiency of the photo diode ($98\%$), the mode matching efficiency ($98\%$), and the clearance of  HD ($96\%$). So, our reconstructed CMs are all corrected with $90\%$ detection efficiency. 

All Gaussian properties can be determined from the CM defined in Eq.~(\ref{cm}). The variances $\Delta^{2}\hat{x}_{A(B)}$ and $\Delta^{2}\hat{p}_{A(B)}$ are obtained by calculating the variances of the measured quadrature values $x_{A(B)}$ and $p_{A(B)}$. The variances of the cross correlations are obtained by $\Delta^{2}(\hat{x}_A\hat{x}_B)=[\Delta^{2}\hat{x}_A+\Delta^{2}\hat{x}_B-\Delta^{2}(\hat{x}_A-\hat{x}_B)]/2$ and  $\Delta^{2}(\hat{p}_A\hat{p}_B)=[\Delta^{2}\hat{p}_A+\Delta^{2}\hat{p}_B-\Delta^{2}(\hat{p}_A-\hat{p}_B)]/2$, where $\Delta^{2}(\hat{x}_A-\hat{x}_B)/2$ and $\Delta^{2}(\hat{p}_A-\hat{p}_B)/2$ are obtained from the simultaneously measured amplitude and phase quadratures of two modes.

To generate two sets of entangled Gaussian states with different squeezing levels, we inject $30$ mW and $50$ mW pump beams into the NOPA respectively. In the case of $30$ mW pump power, the entangled state with lower squeezing ($-1.302/+1.407$ dB) but higher purity is prepared. This can be derived by measuring CM of the prepared entangled state. For instance, for the transmission efficiencies $\eta_A=\eta_B=0.9$, the experimentally measured CM is 
\begin{equation}
\left(\begin{array}{cccc}
1.056\pm0.004 & 0 & -0.287\pm0.004 & 0 \\ 
0 & 1.055\pm0.003 & 0 & 0.287\pm0.002 \\ 
-0.287\pm0.004 & 0 &1.056\pm0.002 & 0 \\ 
0 & 0.287\pm0.002 & 0 & 1.056\pm0.004 
\end{array}\right).
\end{equation}
Based on this CM and the values of efficiencies, we can derive the corresponding variances of sum and difference in Eq.~(\ref{cm}) $V_+= 0.74$ and $V_-=1.38$ ($-1.302/+1.407$ dB).

In the case of $50$ mW pump power, the entangled Gaussian state with higher squeezing ($-1.74/+2.08$ dB) but lower purity is prepared. The CM at the transmission efficiencies $\eta_A=\eta_B=0.9$ is 
\begin{equation}
\left(\begin{array}{cccc}
1.130\pm0.002 & 0 & -0.421\pm0.004 & 0 \\ 
0 & 1.127\pm0.004 & 0 & 0.420\pm0.004 \\ 
-0.421\pm0.004 & 0 &1.127\pm0.004 & 0 \\ 
0 & 0.420\pm0.004 & 0 & 1.128\pm0.004
\end{array}\right).
\end{equation}
Similarly, we have $V_+= 0.67$ and $V_-= 1.61$ ($-1.74/+2.08$ dB). 

\subsection*{C. Remote preparation of the Wigner-negative state at Bob's node}

To generate the Wigner-negative state at Bob's node, we need to subtract a single photon from mode $A$. As shown in Fig.~\ref{figExperimentalSchematic}, around $4\%$ of the energy from mode $A$ is tapped by VBS and directed to the superconducting nanowire single-photon detector (SNSPD). An OI with transmission efficiency $\sim 95\%$ is used to avoid the reflection from the Fabry-Perot cavity (FPC). An interference filter (IF) of $0.6$ nm bandwidth together with a FPC of around $400$ MHz bandwidth are used to filter out the non-degenerate modes. A laser shutter (SR475) is used to match the detection period of the SNSPD with the hold time period of NOPA. Then, the subtracted photon is coupled to the optical fiber by a five-axis fiber aligner and detected by the SNSPD with $\sim 70\%$ detection efficiency whose temperature is cooled to $3$ Kelvin. The output signal of the SNSPD driver (Single quantum, Argos410) is used to trigger the OSC. The OSC with sampling rate of $2.5$ GS/s is used to record the electrical signal of HD 2. 

The dark count in our experiment represents the click of the SNSPD when no down-converted light is incident which comes from the insufficient chopping of the locking beam and the back-scattered light from the LO. When the transmission efficiency of mode $A$ is $\eta_A=0.9$, the dark count rate $R_d$ is around 60 Hz, considering the total counting rate $R_t$ of around $3$ kHz when the pump power is $30$ mW,  $\xi\approx98\%$ (calculated by $1-R_d/R_t$). When the pump power is $50$ mW, the total counting rate $R_t$ is around $7$ kHz, leading to $\xi\approx99\%$. In our experiment, the total photon counting rate will decrease with the decrease of transmission efficiency $\eta_A$, and so will the dark count rate, thus the value of $\xi$ is almost unchanged as it depends on the ratio between them. However, the generation rate of non-Gaussian state at Bob's node is decreased, e.g., for the case with $30$ mW pump power, the generation rate is decreased from $\sim$3 kHz to $\sim$500 Hz when $\eta_A$ decreases from $0.9$ to $0.3$. 
\begin{figure*}[tb]
\centering
\includegraphics[width=0.8\linewidth]{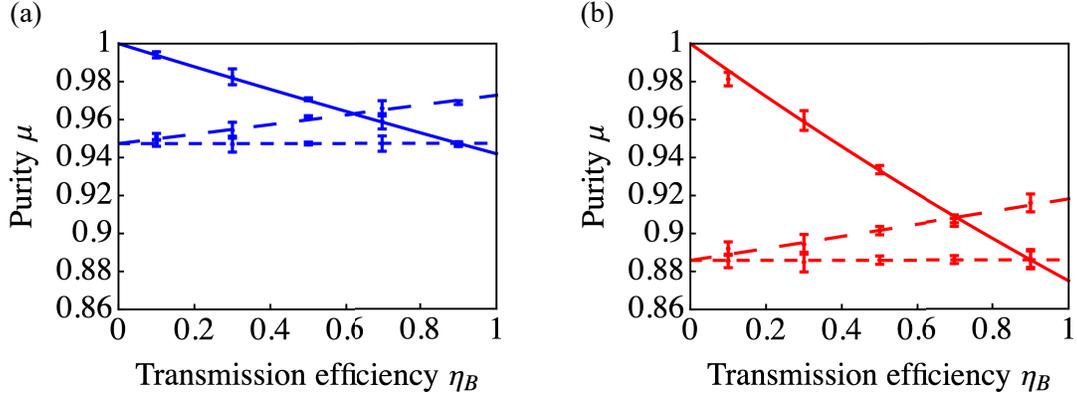}
\caption{Local purities $\mu_{A}$ (short-dashed lines) and $\mu_{B}$ (solid lines) and global purity $\mu_{AB}$ (long-dashed lines) of the initial Gaussian EPR state as functions of transmission efficiency of mode $B$ for two sets of squeezing levels (a) $-1.302/+1.407\text{dB}$ and (b) $-1.74/+2.08\text{dB}$. The parameters are the same as those in Fig. 3 in the main text.}
\label{figPurities}
\end{figure*}

\section*{Appendix III: Relation between Wigner negativity and purities of initial EPR state}

\begin{figure*}[tb]
\centering
\includegraphics[width=1\linewidth]{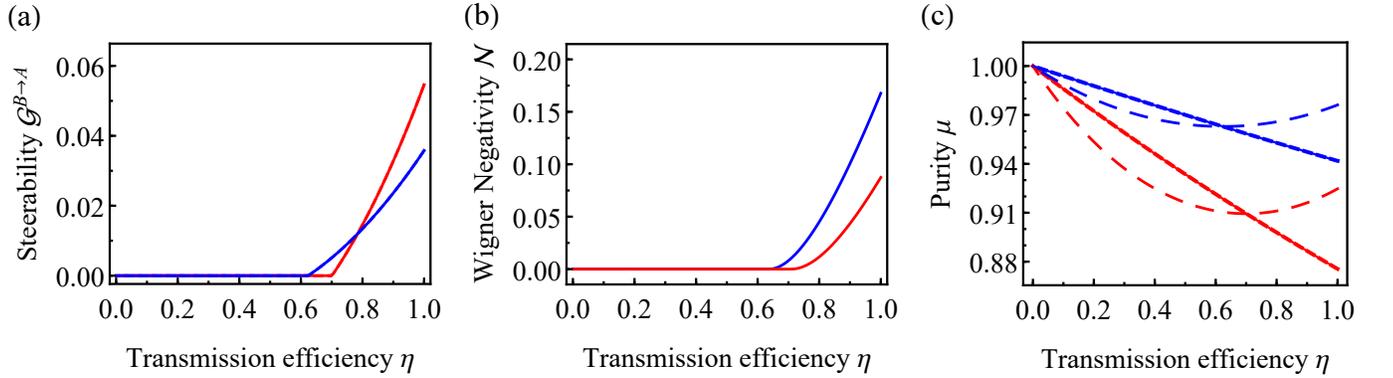}
\caption{(a) Steerability $\mathcal{G}^{B\rightarrow A}$ of the initial EPR state as a function of transmission efficiency $\eta_A=\eta_B=\eta$. (b) The Wigner negativity remotely created in steering mode $B$ by a single-photon subtraction on the steered mode $A$. (c) Local purities $\mu_{A}$ (short-dashed lines), $\mu_{B}$ (solid lines), and global purity $\mu_{AB}$ (long-dashed lines) of the initial EPR state. In order to facilitate comparison, the same two sets of squeezing levels are used: $-1.74/+2.08\text{dB}$ (red lines) and $-1.302/+1.407\text{dB}$ (blue lines). }
\label{figSameEta}
\end{figure*}
In the main text Fig. 3, we have shown that the generated Wigner negativity in mode $B$ only exists when mode $B$ can steer mode $A$, while the amount of negativity is not determined by the Gaussian steerability. As shown in Eq. (\ref{eqWNPurity}), the negativity is quantified by the purities of the prepared EPR state~\cite{XiangarXiv}. To complement the results, Fig. \ref{figPurities} shows the purities of the prepared EPR state versus the transmission efficiency $\eta_B$ corresponding to Fig. 3.

In this case, the transmission efficiency of Alice's channel is fixed at $\eta_A=0.9$. As transmission efficiency $\eta_B$ increases, the local purity of mode $A$ ($\mu_A$) remains invariant because the loss in Bob's channel does not affect Alice's state. While both the local purity of mode $B$ ($\mu_{B}$) and the global purity ($\mu_{AB}$) change with Bob's transmission efficiency, and they become more sensitive to $\eta_B$ with higher squeezing levels. Since the remotely created Wigner negativity is determined by the purities, it explains why the negativity created by higher squeezing level (red) is less robust to channel loss, indicated by Fig. 3(b) in the main text.

Then we consider another scheme where both Alice and Bob's channels suffer from transmission losses. For simplicity, we suppose the degrees of two transmission efficiencies are the same $\eta_A=\eta_B=\eta$. As shown in Fig.~\ref{figSameEta}(a), Gaussian steerability $\mathcal{G}^{B\rightarrow A}>0$ requires the transmission efficiency $\eta>0.701$ in the case with higher squeezing but lower purity (red line), and requires  $\eta>0.623$ in the case with lower squeezing but higher purity (blue line). The Wigner negativity shown in Fig.~\ref{figSameEta}(b) appears when $\eta>0.709$ and $\eta>0.637$ for two cases, respectively, which are slightly higher due to the imperfect photon subtraction. Similar to the case shown in Fig.~3 in main text, the EPR state produced by a higher level of squeezing possesses stronger Gaussian steerability as expected, but the Wigner negativity created in the steering mode $B$ is lower instead. This means that the Gaussian steerability is not directly related to the amount of Wigner negativity. As indicated in Eq.~(\ref{eqWNPurity}), the amount of Wigner negativity depends on the purity of the initially shared EPR state between two modes. Comparing Fig.~\ref{figSameEta}(b) and Fig.~\ref{figSameEta}(c), we find that the case with lower squeezing level but higher purities (blue) remotely generates larger amount of Wigner negativity.

\end{document}